# Rheology and Dynamic Arrest in Colloidal Depletion Gels Mediated by Surface Brush Density

Ziye Zhuang[1], Robert A. Campbell[2], Safa Jamali[2,3], and Ali Mohraz[1*]

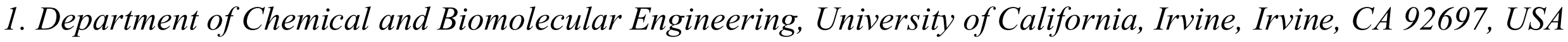

*1. Department of Chemical and Biomolecular Engineering, University of California, Irvine, Irvine, CA 92697, USA*

*2. Department of Mechanical and Industrial Engineering, Northeastern University, Boston, MA 02115, USA*

*3. Department of Chemical Engineering, Northeastern University, Boston, MA 02115, USA*

**Corresponding author: mohraz@uci.edu*

**Acknowledgments**: The authors gratefully acknowledge Dr. Jacinta C. Conrad and Mariah J. Gallegos for helpful discussions and technical consultation on colloidal particle synthesis during the initial development of this work.

**Funding:** Financial support for this study was provided to S.J. and A.M. by the National Science Foundation (PMP-2025613). Z.Z. and A.M. acknowledge the use of facilities and instrumentation at the UC Irvine Materials Research Institute (IMRI), which is supported in part by the National Science Foundation through the UC Irvine Materials Research Science and Engineering Center (DMR-2011967).

## Abstract

We use the density of surface-grafted polymers as a geometry-preserving control parameter for tuning the rheology of colloidal depletion gels. Reducing brush density accelerates gelation and produces gels with higher plateau storage modulus and yield stress. This mechanical enhancement is not accompanied by increased local densification; low-brush networks exhibit lower average contact number and reduced spatial heterogeneity while displaying stronger elastic responses than their high-brush counterparts. Our findings demonstrate a reduced coordination threshold to form elastic nodes in the low-brush gel network. In addition, low-brush gels relax more slowly, accumulate less creep deformation, and exhibit lower effective noise temperatures within the Soft Glassy Rheology framework. These results establish surface-brush density as an experimentally accessible control parameter for colloidal depletion gel rheology with coupled changes in effective attraction, network architecture, and contact kinematics.

## Introduction

Colloidal gels are nonequilibrium soft materials whose macroscopic mechanical and rheological properties emerge from a percolated network of particles connected by attractive interparticle interactions (Weitz and Oliveria 1984; Buscall et al. 1988; Krall and Weitz 1998; Trappe et al. 2001; Studart et al. 2011; Colombo and Del Gado 2014; Landrum et al. 2016). Their elasticity, yielding, and stress relaxation are governed primarily by the constraints that regulate particle motion within the network (Cipelletti et al. 2000; Schweizer and Saltzman 2003; Kobelev and Schweizer 2005; Rajaram and Mohraz 2010; Mewis and Wagner 2013; Kim et al. 2014). These constraints can arise from crowding-induced cages at high particle volume fraction, attractive interparticle forces that result in clustering and limit relative particle displacements, or additional contact-level interactions that suppress rotational and bending degrees of freedom (Yanez et al. 1999; Zaccarelli 2007; Zaccone et al. 2009; Whitaker et al. 2019; Mangal et al. 2024; Nabizadeh et al. 2024). Identifying experimentally accessible ways to tune these particle-level constraints is therefore important for controlling gelation, elasticity, yielding, and long-time rheological response in colloidal gels.

A widely studied class of colloidal gels is formed through short-range attractions induced by nonadsorbing polymers, termed depletion interactions (Asakura and Oosawa 1958; Ilett et al. 1995). In these systems, the polymer concentration provides a direct way to tune interparticle attraction strength. The resulting disordered, space-spanning networks exhibit rheological properties that depend sensitively on particle concentration, interaction strength (via the polymer concentration), and network architecture (Ramakrishnan et al. 2004; Huh et al. 2007; Lu et al. 2008; Laurati et al. 2011; Kim et al. 2015). Their rich rheological behavior, spanning gelation, aging, yielding, stress relaxation, and creep, has motivated efforts to better understand and model how microscopic structure and interparticle interactions govern macroscopic rheology (Wolthers et al. 1997; Zaccone et al. 2009; Sprakel et al. 2011; Hsiao et al. 2014; Varga and Swan 2015; Johnson et al. 2019; Mangal et al. 2023).

One relevant framework for describing long-time rheological dynamics in disordered arrested materials is the Soft Glassy Rheology (SGR) framework, which represents relaxation and flow in terms of activated rearrangements governed by an effective noise temperature (Sollich et al. 1997; Fielding et al. 2000). Although SGR was not developed specifically for depletion-induced colloidal gels, it provides a useful phenomenological approach for analyzing time-dependent rheology in colloidal glasses and gels and enables quantitative comparison of the extent of dynamic arrest across different material conditions (Yin and Solomon 2008; Radhakrishnan et al. 2017).

For spherical colloids interacting through depletion forces, interparticle interactions are typically treated as centrosymmetric because they arise from the overlap of polymer-excluded shells surrounding neighboring particles, whose geometry follows that of the primary particles (Louis et al. 2002; Yang et al. 2006). Within this central-force description, mechanical stability of colloidal depletion gels can be rationalized using Maxwell-type constraint counting, which relates rigidity to the balance between particle degrees of freedom and the number of independent constraints imposed by particle contacts (Maxwell 1864; Hsiao et al. 2012; Rocklin et al. 2021). For frictionless spherical particles with purely central interactions in three dimensions, this counting gives the familiar isostatic contact criterion of $Z = 6$, where $Z$ denotes the number of interparticle contacts. However, when additional contact-level constraints such as frictional, bending, or rotational resistance are present, rigidity can emerge at contact numbers below the central-force threshold (Pantina and Furst 2005; Wang and Swan 2019; Immink et al.

2020; Nguyen et al. 2020). Such particle-level constraints can significantly alter how rigidity and relaxation dynamics in colloidal gels are related to particle contact number.

Non-central interparticle interactions have been explored in systems with patchy particles, surface roughness, or anisotropic geometries, where directional bonding or frictional contacts introduce constraints on particle dynamics (Del Gado and Kob 2010; Hsiao et al. 2017; Sciortino and Zaccarelli 2017; Scherrer et al. 2025). These studies demonstrate that non-central forces can alter gelation pathways, network structure, and bulk rheology of colloidal gels (Bianchi et al. 2006; Hsu et al. 2018; Müller et al. 2023; Dias et al. 2025; Palombo et al. 2025). Here, we introduce an alternative and experimentally simple route for imposing such constraints without changing particle geometry or introducing significant surface roughness. This approach relies on reducing the grafting density of the stabilizing polymer brushes on the particle surface.

Polymer-grafted surface brushes are commonly used to stabilize colloids against aggregation through steric or electro-steric repulsion (Pincus 1991; Akpinar et al. 2016). In our previous work (Zhuang et al. 2026), we showed that reducing the density of this stabilizing layer provides a facile route to activate non-central constraints in colloidal depletion gels while preserving spherical particle geometry and the underlying depletion mechanism. Specifically, this reduction enables partial interpenetration of neighboring brush layers, introducing an effective bending rigidity at particle contacts without requiring shape anisotropy, patchiness, or surface roughness. At a low particle volume fraction ($\phi$ = 0.10), this modest change in brush density produced distinct gel morphologies: low-brush particles formed more open, tenuous networks, whereas high-brush particles formed more compact, locally densified structures.

Building on this framework, the present study examines how the same brush-density reduction influences the rheological response of depletion-induced colloidal gels at a more concentrated particle volume fraction ($\phi$ = 0.20). In this regime, crowding can reduce the morphological contrast between the structures but amplify differences in their rheology in both the linear and nonlinear regimes. Using core-shell particles with tunable brush density, we directly compare high-brush and low-brush systems across a range of depletant concentrations while holding particle size, shape, and volume fraction fixed. We show that reducing brush density accelerates gelation, increases stiffness and yield stress, and suppresses long-time stress relaxation dynamics despite reduced coordination. By combining quantitative microstructural analysis with rheological measurements and theoretical analysis in the context of the SGR model, we demonstrate that a simple reduction in brush density enables rigidity below the classical centrosymmetric isostaticity threshold and drives the system into deeper dynamic arrest. Together, these results show for the first time that modulating surface brush density can provide a route for controlling rheology and dynamic arrest in colloidal depletion gels without altering particle geometry. Note that because reducing surface-brush density can alter both the effective interparticle potential and the kinematic constraints at particle contacts, the resulting rheological differences should be interpreted as the combined consequences of interaction strength, network architecture, and contact mechanics. The coordination analysis presented below therefore tests whether the observations are consistent with additional angular constraints; it does not independently isolate or directly measure those constraints.

## Materials and Methods

### Core-shell Particle Synthesis

Monodisperse core-shell colloidal particles with mean diameter $d_p$ = 1.06 ± 0.027 µm were synthesized following modified procedures based on previous work (Kodger et al. 2015;

Park et al. 2018). The synthesis consisted of three steps: (i) preparation of fluorescent polymer cores, (ii) growth of a non-fluorescent shell, and (iii) grafting of charged surface polymer brushes for colloidal stabilization. All chemicals were purchased from Sigma Aldrich unless otherwise noted.

**Core synthesis.** Fluorescent polymer cores were prepared using a 45:55 volumetric ratio of 2,2,2-trifluoroethyl methacrylate (TFEMA, 3.21 mL) and *tert*-butyl methacrylate (tBMA, 3.93 mL), with ethylene glycol dimethacrylate (EGDMA, 0.147 mL) added as a crosslinker. Fluorescence was incorporated via coumarin-methacrylate (2.5 mL), synthesized following the procedure of Kodger et al. 2015. Initiation was carried out using 2,2′-azobis(2-methylpropionitrile) (AIBN, 0.077 g), with 3-sulfopropyl methacrylate potassium salt (0.077 g) and 2-(2-bromoisobutyryloxy) ethyl acrylate (0.143 mL, synthesized following Kodger et al. 2015) added for surface functionality. All reagents were dispersed in a methanol-water mixture (170.2 mL methanol, 37.1 mL water), and polymerization proceeded at 80 °C under reflux for 5 h with continuous stirring.

**Shell growth.** A non-fluorescent shell was subsequently grown using the same TFEMA-tBMA ratio (2.70 mL and 3.30 mL, respectively), with poly(vinylpyrrolidone) (PVP, 2.44 g) serving as a steric stabilizer. The fluorescent cores (15 mL, 20 vol% in methanol-water) and 2-(2-bromoisobutyryloxy) ethyl acrylate (0.36 mL) were added to a methanol-water mixture (155.4 mL methanol, 11.9 mL water). Polymerization was initiated with AIBN (0.076 g) at 55 °C and allowed to proceed for 16 h.

**Surface functionalization.** Charged stabilizer brushes were grafted through copolymerization of 2-acrylamido-2-methyl-1-propanesulfonic acid sodium salt (SPAm, 2.52 mL) and *N,N*-dimethylacrylamide (DMA, 0.678 mL) in the presence of copper(I) chloride (0.028 g), copper(II) chloride (0.035 g), and hexamethyltriethylenetetramine (0.151 mL). The core-shell suspension (15 mL, 25 vol% in methanol-water) and a sacrificial initiator (PEGini, 0.289 mL, synthesized following Kodger et al. 2015) were added to a methanol (6.74 mL)-water (6.68 mL) mixture. Because the exact number of accessible surface ATRP initiation sites is not independently measurable in this system, PEGini was added in large excess to regulate the overall chain-growth statistics. The reaction proceeded for 6 hours at room temperature, yielding colloidal particles coated with dense charged polymer brushes ("high-brush" particles).

**Low-brush particles.** To prepare particles with reduced surface brush density, we modified both the shell growth and surface functionalization steps. During shell growth, the amount of 2-(2-bromoisobutyryloxy) ethyl acrylate was decreased by 25% (from 0.36 mL to 0.27 mL) to reduce the density of brush initiation sites on the particle surface. The subsequent surface functionalization was carried out identically to the high-brush case but with a proportional 25% reduction in DMA, SPAm, and PEGini concentrations (1.89 mL SPAm, 0.509 mL DMA, and 0.217 mL PEGini). By approximately preserving the monomer-to-initiator ratio, the average brush length is expected to remain comparable to the high-brush system while reducing the overall relative brush density. These modifications consistently produced colloids with reduced surface brush density ("low-brush" particles).

**Preparation of Colloidal Gels**

A 75/25 wt% glycerol-water mixture was used as the solvent to provide near refractive-index and density matching with the synthesized particles. Non-adsorbing polyacrylamide (PAM, $M_n = 150{,}000$) and sodium chloride (NaCl) were dissolved in this solvent to prepare concentrated stock solutions of polymer depletant and salt.

For each sample, appropriate volumes of stock solvent, PAM solution, NaCl solution, and particle suspension were mixed in a vial to yield a final particle volume fraction of $\phi$ = 0.20, a NaCl concentration of 10 mM, and a PAM concentration of 1, 2, 3, 4, or 6 mg/mL. Importantly, the addition of salt alone did not visibly affect colloidal stability in either brush system, as no obvious aggregation was observed (see Figure S1 in Online Resource 1 for details). Each suspension was vortexed thoroughly to ensure complete homogenization and uniform particle dispersion prior to all measurements.

**Confocal Microscopy and Structural Measurements**

Confocal imaging was performed using a laser scanning microscope (FLUOVIEW FV3000, Olympus) equipped with a 40× oil-immersion objective. Samples were transferred into a custom-built glass imaging chamber and allowed to gel undisturbed for 2 h prior to imaging. To capture the three-dimensional (3D) microstructure, a stack of two-dimensional (2D) images was acquired at vertical (z) intervals of 0.25 μm, beginning 5 μm above the coverslip and spanning a total depth of 30 μm. The imaging field of view was 50 × 50 μm in the x-y plane, and three independent z-stacks were collected at different locations for each sample.

The gel microstructures were quantified using Python-based algorithms of computational geometry as follows. A particle tracking algorithm (TrackPy by Allan et al. 2025) was first used to locate particle centers with subpixel accuracy. Particles were identified as nearest neighbors if their interparticle separation fell within the defined search distance criteria (details provided in Online Resource 1). Custom analysis scripts were then used to compute structural metrics, including particle contact numbers, Voronoi volumes, and network connectivity. Particle contact networks were constructed from nearest-neighbor pairs. The largest connected component (LCC) was defined as the largest group of particles connected to one another through nearest-neighbor contacts, and the fraction of total particles belonging to the LCC was used to quantify network connectivity. To assess the fraction of particles most prominently responsible for the sample's load-bearing characteristics, we introduced a coordination-defined load-bearing particle fraction based on contact-number thresholds. Two criteria were considered: Maxwell's classical isostatic criterion (Maxwell 1864; Hsiao et al. 2012) for nodes with centrosymmetric interactions ($Z \geq 6$), and a reduced threshold ($Z \geq 3$) motivated by our prior work (Zhuang et al. 2026), in which brush-mediated non-central interactions are hypothesized to suppress rotational degrees of freedom and enable mechanical stability at lower coordination numbers. In all cases, load-bearing particle fractions were computed with respect to the total number of particles within the LCC.

**Rheology**

Rheological measurements were conducted using a Discovery Hybrid Rheometer (HR-3, TA Instruments) to characterize the viscoelastic properties and gelation dynamics of the colloidal suspensions at a constant temperature of 20 °C. All experiments were performed using a 40 mm parallel-plate geometry. Samples were loaded onto the stationary lower plate, and the gap was set to 500 μm. Excess material was carefully trimmed, and a custom solvent-trapping cap was used to minimize evaporation during long-duration measurements.

Time-sweep measurements were performed to monitor gelation and structural evolution by tracking both the storage ($G'$) and loss ($G''$) moduli over 120 min at a strain amplitude of 0.5% and an angular frequency of 1 rad/s. Prior to each measurement, samples were pre-sheared at a shear rate of 50 $s^{-1}$ for 60 s to eliminate pre-existing structures and ensure a reproducible initial state.

Following the time sweeps, strain-sweep measurements were carried out to probe the samples' nonlinear mechanical properties and yielding behavior. The strain amplitude was increased from 0.01% to 500% at 1 rad/s, and the yield point was identified as the crossover of $G'$ and $G''$.

Stress-relaxation measurements were performed by applying a step strain of 0.5%, selected to remain within the linear viscoelastic regime, and monitoring the decay of the relaxation modulus $G(t)$ over 1800 s. Creep measurements were conducted by applying a constant shear stress of 0.0105 Pa and recording the resulting compliance $J(t)$ over 1800 s. This stress was chosen to remain within the linear viscoelastic regime for all gelled samples, enabling long-time compliance measurements without inducing yielding. For both relaxation and creep tests, samples were first pre-sheared and then allowed to age for 7200 s prior to measurement.

**Soft Glassy Rheology (SGR) Model Fitting**

The stress relaxation and creep behavior of the gels were analyzed within the Soft Glassy Rheology (SGR) framework, which describes the rheology of disordered, arrested soft materials in terms of activated yielding events over an energy landscape (Sollich et al. 1997; Fielding et al. 2000). Although originally developed for glassy soft materials, the applicability of this framework to colloidal gels has been established in prior studies (Yin and Solomon 2008). In this model, the material is represented as an ensemble of elements trapped in energy wells, which yield through activated rearrangements governed by an effective noise temperature $x$. Lower values of $x$ correspond to more deeply arrested, glassy dynamics, while larger values indicate increasingly fluid-like behavior.

**Stress relaxation**. Stress relaxation following a small step strain was first analyzed to extract the effective noise temperature. The relaxation modulus $G(t)$ was fitted using the analytical expression derived within the SGR framework and later applied to colloidal gels (Fielding et al. 2000; Yin and Solomon 2008), which accounts for aging via a waiting time, $t_w$:

$$\frac{G\left(\frac{t-t_w}{\tau_0}, \frac{t_w}{\tau_0}\right)}{G\left(0, \frac{t_w}{\tau_0}\right)} \approx 1 - \frac{\Gamma(x)\left[\left(\frac{t-t_w}{\tau_0}\right)^{1-x} - 1\right]}{\Gamma^2(x)\Gamma(2-x)\left[\left(\frac{t_w}{\tau_0}\right)^{1-x} - 1\right]} \quad (1)$$

The experimentally measured $G(t)$ was normalized by its initial value and fitted using nonlinear least-squares regression. The effective noise temperature $x$ and microscopic attempt time $\tau_0$ were treated as fitting parameters, while all other quantities were fixed by experimental conditions. This procedure yielded robust values of $x$ for all samples, providing a quantitative measure of dynamic arrest across all tested conditions.

**Creep deformation**. To further assess the consistency of the SGR predictions for our materials, we analyzed the linear creep response under a constant applied stress using the same framework. In the linear regime, the creep compliance $J(t-t_w, t_w)$ is governed by the SGR constitutive equation (Fielding et al. 2000):

$$J(t-t_w, t_w) = 1 + \int_{t_w}^{t} J(t'-t_w, t_w) Y(t') G_p(t-t') dt' \quad (2)$$

where $Y(t')$ is the time-dependent yielding rate, and $G_p(t-t')$ is the survival function that represents the retained elastic contribution from elements that remain trapped between times $t'$ and $t$.

Because this expression does not provide a general closed-form solution, different asymptotic forms were used depending on the value of $x$. The appropriate regime for each sample was determined based on the value of $x$ obtained independently from stress relaxation:

- For $x < 1$ (glassy regime), the full integral equation was solved numerically using asymptotic expressions for $Y(t)$ and $G_\rho(t - t')$ derived within the SGR framework (Fielding et al. 2000). The explicit expressions used are provided in Online Resource 1, S2.
- For $1 < x < 2$ (fluid-like regime), the short-time asymptotic solution provided by Fielding and coworkers (Fielding et al. 2000) was used:

$$J(t - t_w, t_w) \sim \frac{(t - t_w)^{x-1}}{\Gamma^2(x)\Gamma(2 - x)} \quad (3)$$

The SGR model yields a dimensionless creep compliance, $J_{SGR}(t)$. To compare this prediction with the experimentally measured compliance, the dimensionless model output was scaled using:

$$J_{fit}(t) = A\, J_{SGR}(t) \quad (4)$$

where $J_{fit}(t)$ is the dimensional compliance used to fit the experimental data and $A$ is a sample-specific compliance scale factor with units of $Pa^{-1}$. For creep fitting, both $x$ and $A$ were treated as adjustable parameters for each sample. The resulting creep-derived values of $x$ were then compared with those obtained independently from stress relaxation to better assess the applicability of the SGR framework for describing the rheology of our materials.

## Results and Discussion

Figure 1 shows representative scanning electron microscopy (SEM) images of the high-brush and low-brush colloids used in this study, together with schematics illustrating the relative difference in surface brush density. For consistency, throughout the paper results associated with high-brush and low-brush particles are presented in shades of green and purple, respectively. The two particle populations are geometrically indistinguishable in size and shape, confirming that reducing brush density does not introduce measurable changes to the primary particle geometry.

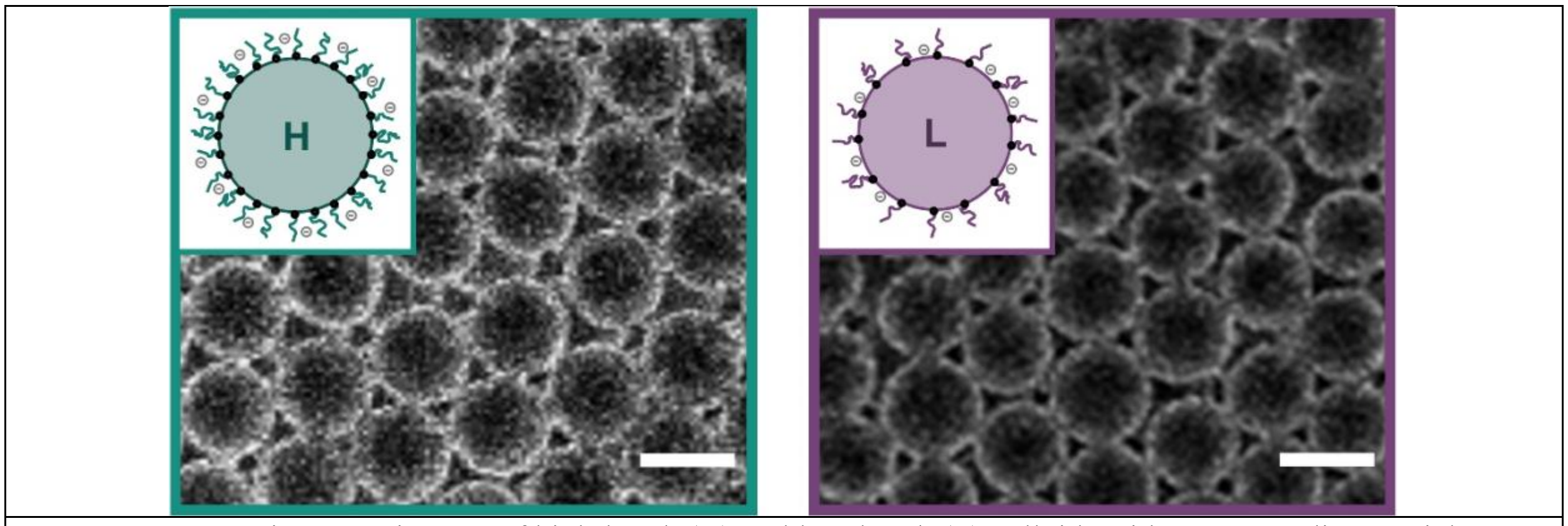

**Fig. 1** Representative SEM images of high-brush (H) and low-brush (L) colloids with corresponding particle schematics shown as insets. Despite differences in surface brush architecture, the particles remain geometrically indistinguishable in size and shape.

## Gelation Kinetics and Elasticity

Figure 2a shows small-amplitude oscillatory shear (SAOS) time sweeps used to examine how reducing brush density influences gel formation and the evolution of mechanical properties. Across all concentrations of the polymer depletant (PAM) tested, both (low- and high-brush) systems exhibit the characteristic evolution of colloidal depletion gels, with a rapid initial increase in elasticity followed by a plateau in the zero-shear elastic modulus, $G_0'$, at long times. As expected, increasing depletant concentration leads to a systematic increase in $G_0'$ for both systems, which is attributed to an increase in the interparticle attraction strength.

Despite these shared trends, brush density produces a clear separation between the two systems in both gelation kinetics and mechanical response. With the exception of the high-brush sample at 1 mg/mL PAM, which remains primarily fluid-like within the experimental window (see Figure S2 in Online Resource 1), all other samples develop a solid-like viscoelastic response ($G' > G''$). At matched depletant concentration, low-brush gels consistently reach higher plateau moduli than their high-brush counterparts. This observation is consistent with our previous work (Zhuang et al. 2026), where lower brush density was shown to allow closer particle approach, thereby enhancing the effective attraction. The enhanced modulus therefore likely reflects coupled changes in effective attraction strength, network architecture, and contact-level kinematics. Because these contributions are not independently varied here, the rheological comparison alone cannot assign the modulus increase uniquely to angular resistance.

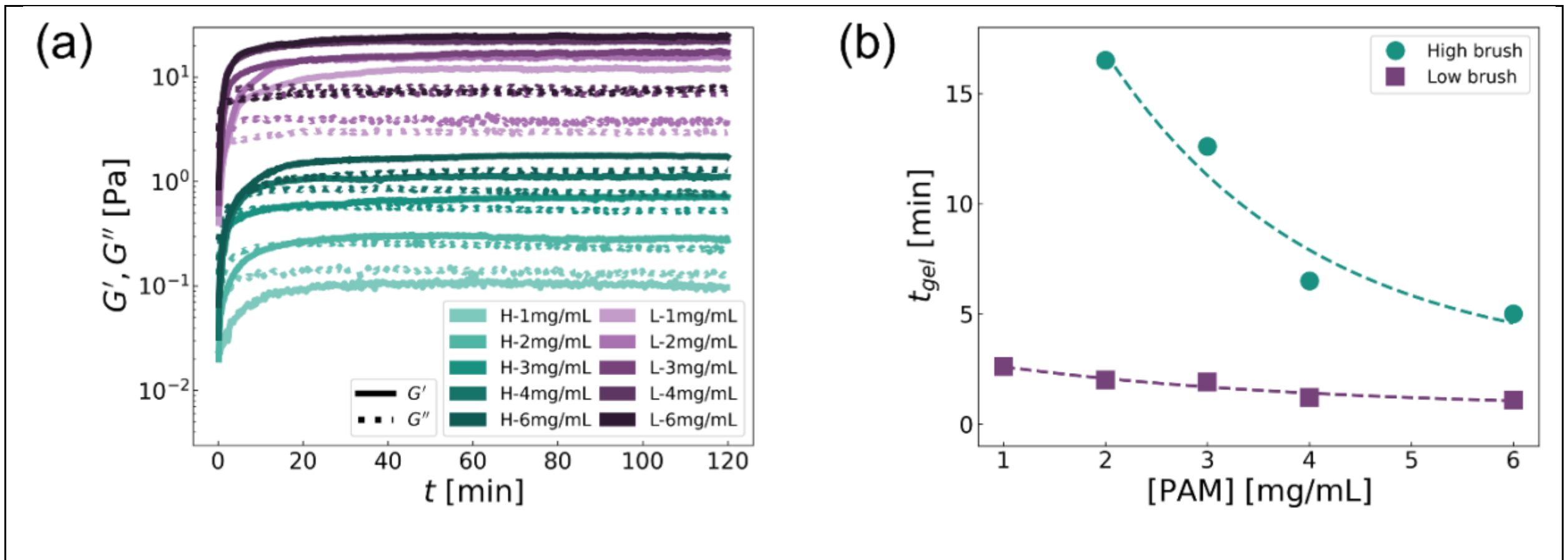

**Fig. 2** Gelation kinetics and linear viscoelastic response of high-brush (H) and low-brush (L) colloidal samples. **(a)** Small-amplitude oscillatory shear (SAOS) time sweeps showing the storage modulus, $G'$, and loss modulus, $G''$, across PAM concentrations of 1–6 mg/mL. Measurements were performed for 120 min at 0.5% strain and 1 rad/s following pre-shear. All samples except the high-brush system at 1 mg/mL developed a solid-like viscoelastic response, defined by $G' > G''$, within the measurement window. **(b)** Gelation time, $t_{\mathrm{gel}}$, defined as the first $G' = G''$ crossover extracted from the time sweeps, plotted as a function of PAM concentration, [PAM]. Low-brush systems gel more rapidly than high-brush systems at matched depletant concentrations. Dashed lines are guides to the eye.

To quantify the kinetics of network formation, we define the gelation time, $t_{\mathrm{gel}}$, as the crossover point where $G' = G''$, and extract this value for each sample from the time sweep results. Figure 2b shows the extracted $t_{\mathrm{gel}}$ as a function of depletant concentration for the low- and high-brush particles. Across all gelled conditions, low-brush systems show faster gelation kinetics, with their $t_{\mathrm{gel}}$ values more than fivefold shorter than those of high-brush particles at low depletant concentrations. This acceleration suggests that reducing brush density affects the formation and early stabilization of the particle network, rather than simply increasing the stiffness of a network after it has formed.

## Network Morphology

To connect these rheological observations to the underlying microstructure, we next examine the arrested network morphology using confocal microscopy, with representative images shown in Figure 3. Once the depletion attraction is sufficiently strong to form a colloidal network (≥ 2 mg/mL PAM), the two systems adopt clearly distinct architectures. As qualitatively seen in Figure 3, high-brush gels form compact domains separated by relatively large voids, whereas low-brush gels form less spatially heterogeneous networks across all polymer concentrations. Importantly, both systems form system-spanning networks at PAM concentrations ≥ 2 mg/mL, as confirmed by near-unity fractions of particles in the largest connected component (LCC), shown in Figure S3 of Online Resource 1. These observations are also consistent with the gelation times of Figure 2b, where the less heterogeneous configurations of the low-brush system result in faster percolation and gelation kinetics.

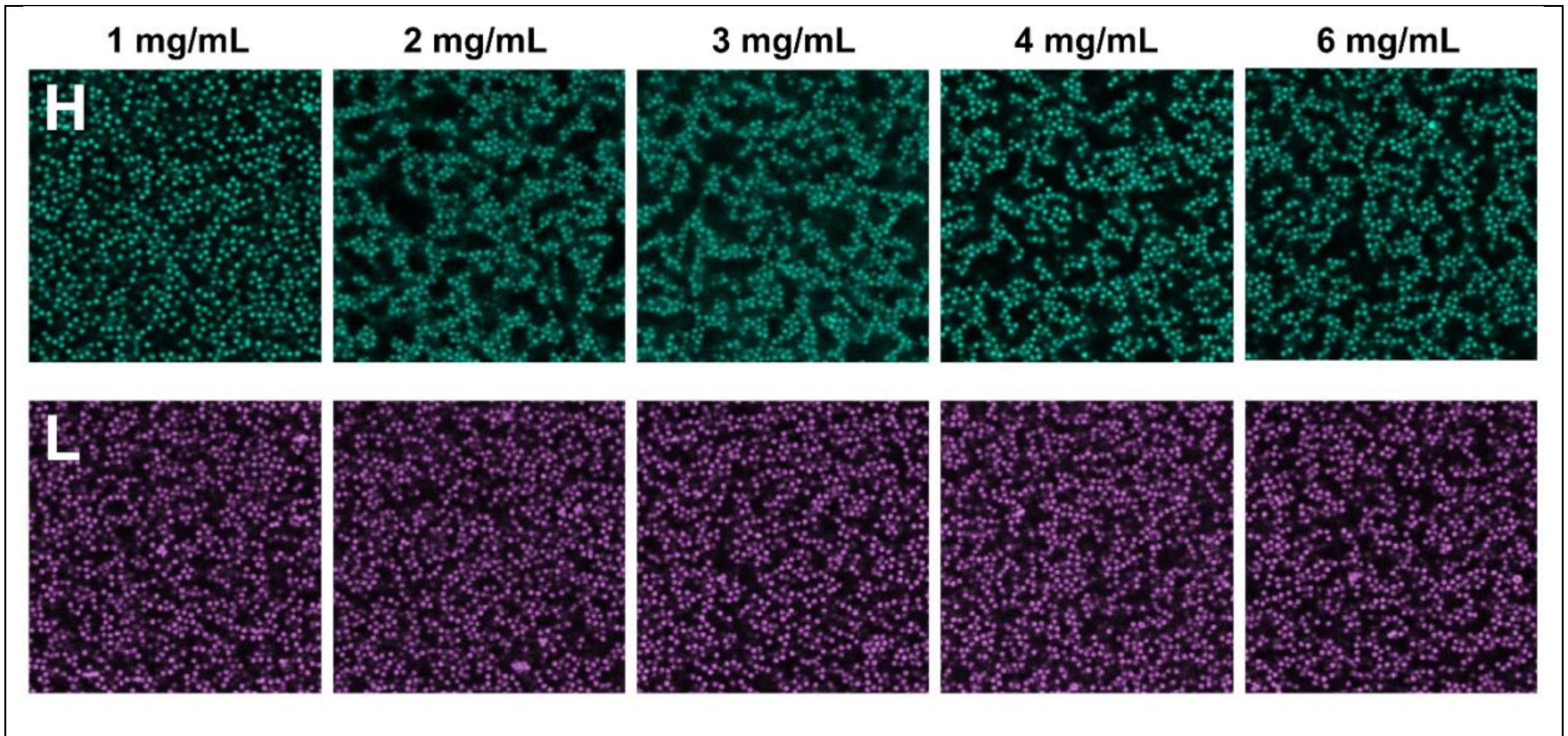

**Fig. 3** Representative confocal images of high-brush (H) and low-brush (L) colloidal samples across PAM concentrations of 1–6 mg/mL, acquired after 2 h of quiescent aging. At PAM concentrations sufficient to form system-spanning networks, high-brush samples exhibit more compact, locally dense structures separated by larger voids, whereas low-brush samples form more open and spatially homogeneous networks.

To quantify the structural differences observed in Figure 3, from the confocal data we compute particle-resolved geometrics such as the number of interparticle contacts, $Z$, and the Voronoi volume associated with each particle, $V_{Voro}$. The distributions of $Z$ and $V_{Voro}$ for a representative condition (3 mg/mL PAM) are shown in Figures 4a and 4b, respectively. The contact-number distributions are similar in shape, although the low-brush system is slightly shifted toward lower coordination values. In contrast, the Voronoi volume distribution shows a more pronounced difference: the high-brush system exhibits a longer high-volume tail, whereas the low-brush system has a narrower distribution with fewer particles associated with large local volumes. This result indicates reduced spatial heterogeneity and fewer large void regions in the low-brush network, consistent with the visual signatures in Figure 3. Contact number and Voronoi volume distributions for all other conditions are presented in Online Resource 1, and show the same signatures as those in Figures 4a and 4b.

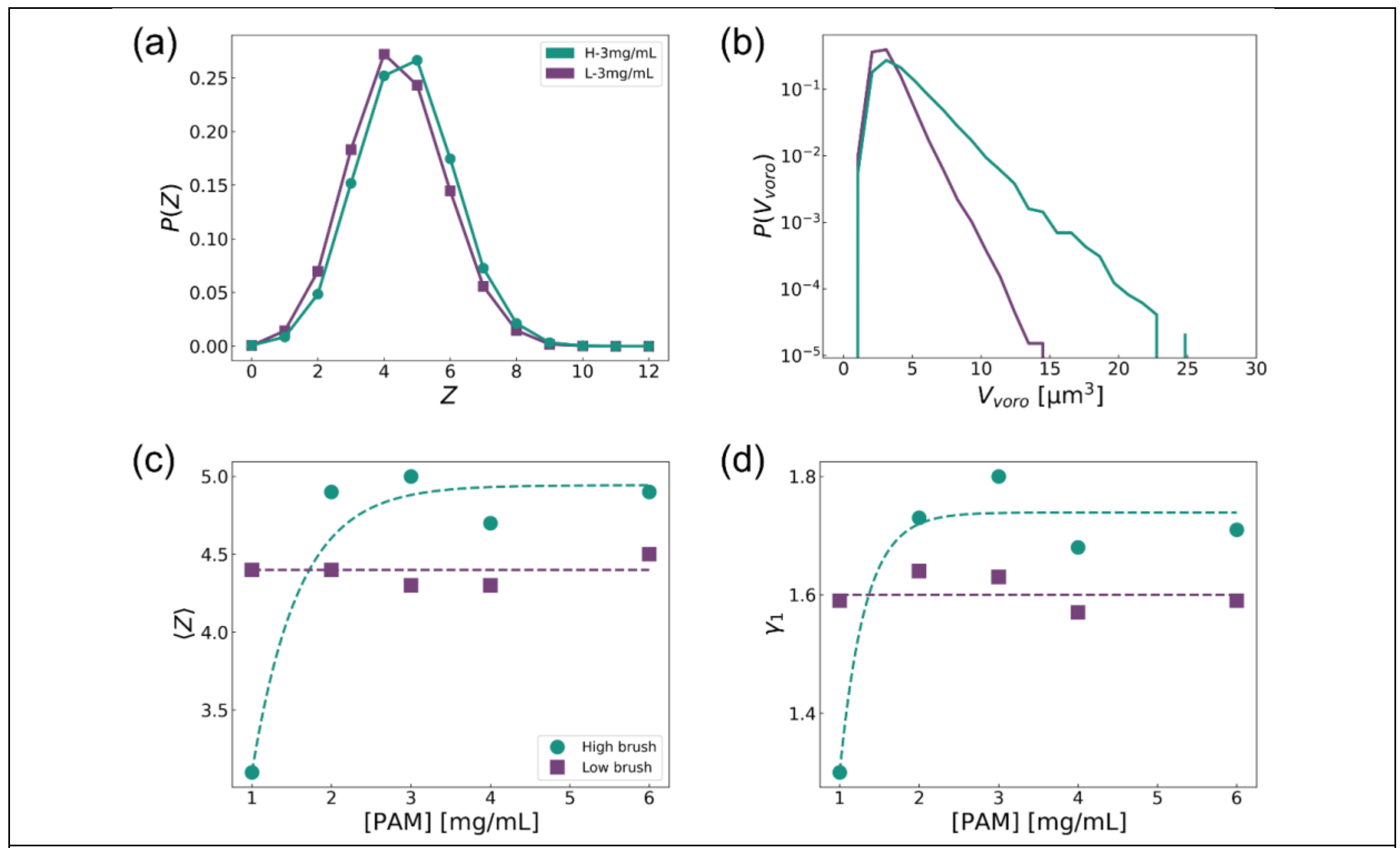

**Fig. 4** Structural characterization of high-brush (H) and low-brush (L) samples. **(a)** Particle contact number distributions and **(b)** Voronoi volume distributions at 3 mg/mL PAM. Distributions for all depletant concentrations are provided in Online Resource 1. **(c)** Average contact number ⟨Z⟩ and **(d)** skewness $\gamma_1$ of the Voronoi volume distribution as functions of PAM concentration, [PAM]. Low-brush systems exhibit lower coordination and reduced spatial heterogeneity. Dashed lines are guides to the eye.

To quantify these representative trends across all depletant concentrations, we next compare the average contact number, ⟨Z⟩, and the skewness of the Voronoi volume distribution, $\gamma_1$, for all conditions tested, as shown in Figures 4c and 4d. The skewness, $\gamma_1$, quantifies the asymmetry of the Voronoi volume distribution, with larger positive values indicating greater spatial heterogeneity in the microstructure. Across all gelled conditions, high-brush samples generally exhibit higher ⟨Z⟩ (Figure 4c), consistent with more locally densified structures. In addition, low-brush samples maintain lower $\gamma_1$ values across all depletant concentrations (Figure 4d), indicating a more uniform distribution of local volumes and reduced mesoscale heterogeneity.

Together, these structural measurements show that the enhanced stiffness of the low-brush gels is not accompanied by increased local densification. Low-brush networks exhibit lower average contact number and reduced spatial heterogeneity (Figures 4c and 4d), yet they reach higher plateau moduli than the corresponding high-brush gels (Figure 2a). Thus, the rheological differences between the two systems cannot be attributed simply to the formation of more contacts or denser local packing. This contrast suggests that in this system, reducing the brush density meaningfully impacts how interparticle contacts contribute to stress support within the network.

**Rigidity and Yielding beyond Central-Force Isostaticity**

Figure 5a shows strain-sweep measurements used to probe the mechanical robustness of the arrested networks after the two-hour time sweep, once $G'$ has reached its plateau value. At small strain amplitudes, all gelled samples exhibit a well-defined linear viscoelastic regime with $G' > G''$ consistent with Figure 2a. As the strain amplitude is increased, $G'$ decreases and eventually crosses $G''$, indicating yielding and loss of elastic dominance. We define the yield strain, $\gamma_y$, as the strain amplitude at the $G' = G''$ crossover, and the yield stress, $\sigma_y$, as the corresponding stress at this point.

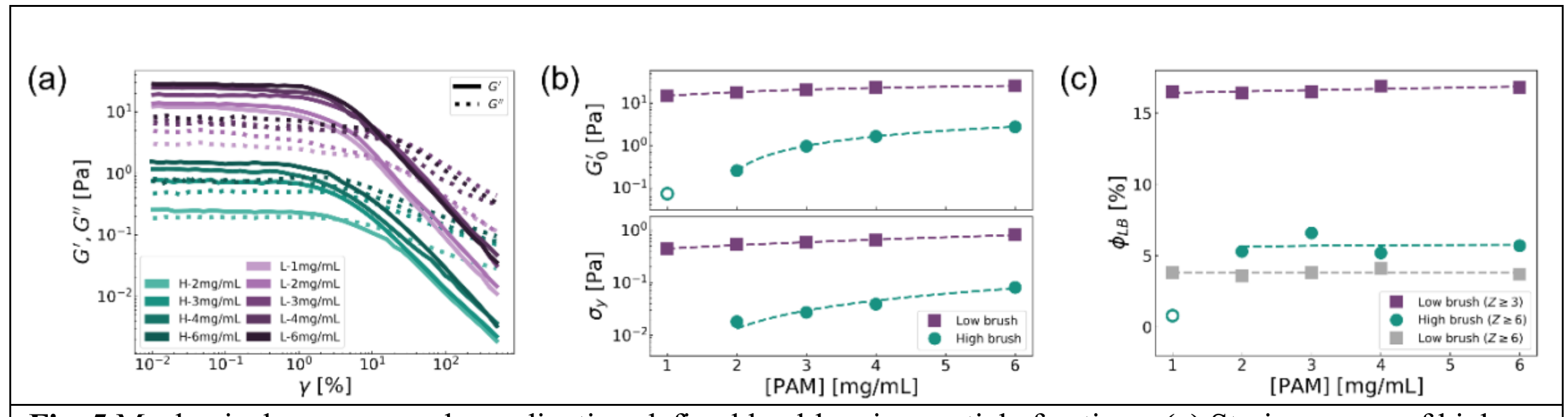

**Fig. 5** Mechanical response and coordination-defined load-bearing particle fractions. **(a)** Strain sweeps of high-brush (H) and low-brush (L) samples performed after 2 h of aging. Gelled samples exhibit a linear viscoelastic regime at small strain followed by yielding at larger strain amplitudes. **(b)** Plateau storage modulus, $G_0'$, and yield stress, $\sigma_y$, as functions of PAM concentration, [PAM], showing consistently higher stiffness and yield stress in low-brush gels. **(c)** Load-bearing particle fraction, $\phi_{LB}$, identified using coordination-based criteria. The classical central-force isostatic threshold, $Z \geq 6$, gives larger load-bearing particle fractions for high-brush gels, whereas a reduced threshold, $Z \geq 3$, for low-brush gels gives trends consistent with their enhanced mechanical response. Dashed lines are guides to the eye. Open symbols denote the non-gelled high-brush sample at 1 mg/mL PAM, which is excluded from analysis.

To further quantify the mechanical response of our samples, we extract the plateau storage modulus, $G_0'$, from the linear viscoelastic regime and $\sigma_y$ from the yield point. Figure 5b shows the dependence of $G_0'$ and $\sigma_y$ on depletant concentration for the low- and high-brush systems. Across all gelled samples, low-brush gels exhibit higher moduli and yield stresses than their high-brush counterparts, confirming that reducing brush density enhances both stiffness and resistance to mechanical failure. Because both systems form system-spanning networks under gelled conditions (recall Figure S3 of Online Resource 1), these differences are not associated with incomplete percolation. Instead, they arise from how load is distributed within the fully connected networks formed by the two particle types.

To connect the observed rheological response to network structure, we estimate the fraction of particles expected to participate in the load-bearing network using coordination-based criteria. We first use the classical central-force isostatic criterion as a reference. For frictionless spherical particles in three dimensions with purely central interactions, Maxwell-type constraint counting predicts an isostatic coordination of $Z = 6$ for mechanical stability (Maxwell 1864; Hsiao et al. 2012). We therefore use $Z \geq 6$ to identify the subset of particles expected to contribute to the load-bearing structure of the network under the central-force criterion, and plot their fraction in Figure 5c. Using this criterion, the high-brush system contains a larger fraction of $Z \geq 6$ particles than the low-brush system across gelled conditions. This trend contrasts with the rheological measurements, where the low-brush gels exhibit higher $G_0'$ and $\sigma_y$. Thus, a particle fraction based only on the central-force coordination threshold does not capture the enhanced mechanical response of the low-brush networks.

We next apply a reduced coordination criterion motivated by the brush-mediated contact mechanics identified in our previous work (Zhuang et al. 2026). When neighboring low-brush particles partially interpenetrate through their brushes, their contacts can resist angular displacement in addition to center-to-center separation. In a simple constraint-counting picture, this angular resistance provides an additional mechanical constraint per contact, reducing the coordination number needed for a particle to contribute to a mechanically stable network. We therefore use $Z \geq 3$ as a reduced coordination criterion for identifying particles that may participate in the load-bearing structure of the low-brush system. As shown in Figure 5c, under this criterion, the low-brush gels exhibit a larger load-bearing particle fraction than the high-brush gels, consistent with their higher plateau modulus and yield stress. Our results indicate that the stronger mechanical response of the low-brush gels is not captured by a load-bearing particle fraction based on the classical central-force threshold alone. Instead, the comparison is more consistent with a picture in which reduced brush density introduces contact-level resistance to angular deformation, allowing particles with lower coordination to contribute to stress support and yielding resistance more effectively than expected from central-force contact counting.

**Stress Relaxation, Creep, and Dynamic Arrest**

Having shown that reduced brush density increases stiffness and yielding resistance despite lower average coordination, we next examine how brush density affects the long-time rheological response of the gels. Stress relaxation and creep provide complementary probes of time-dependent network rearrangement under small deformation: stress relaxation measures how a gel dissipates stress after an imposed strain, whereas creep measures how the network accumulates deformation under a constant applied stress. If reduced brush density changes the mechanics of particle contacts and the load-bearing network, these effects should also be reflected in the rate at which the gel relaxes or deforms over time.

Figure 6 presents the long-time rheological response of the gels measured using two complementary small-deformation protocols: stress relaxation following a 0.5% step strain (Figure 6a) and creep under a constant applied stress of 0.0105 Pa (Figure 6b). In the stress-relaxation measurements, low-brush gels retain higher relaxation moduli, $G(t)$, over the experimental window than their high-brush counterparts. This behavior is consistent with the higher plateau moduli observed in Figure 2a and indicates that the low-brush networks dissipate stress more slowly after deformation. In the creep measurements, low-brush gels exhibit lower compliance, $J(t)$, and slower compliance growth, indicating that they accumulate less deformation under the same applied stress. The solid lines in Figure 6 represent fits to the corresponding stress-relaxation and creep responses using the SGR expressions described in the Materials and Methods section.

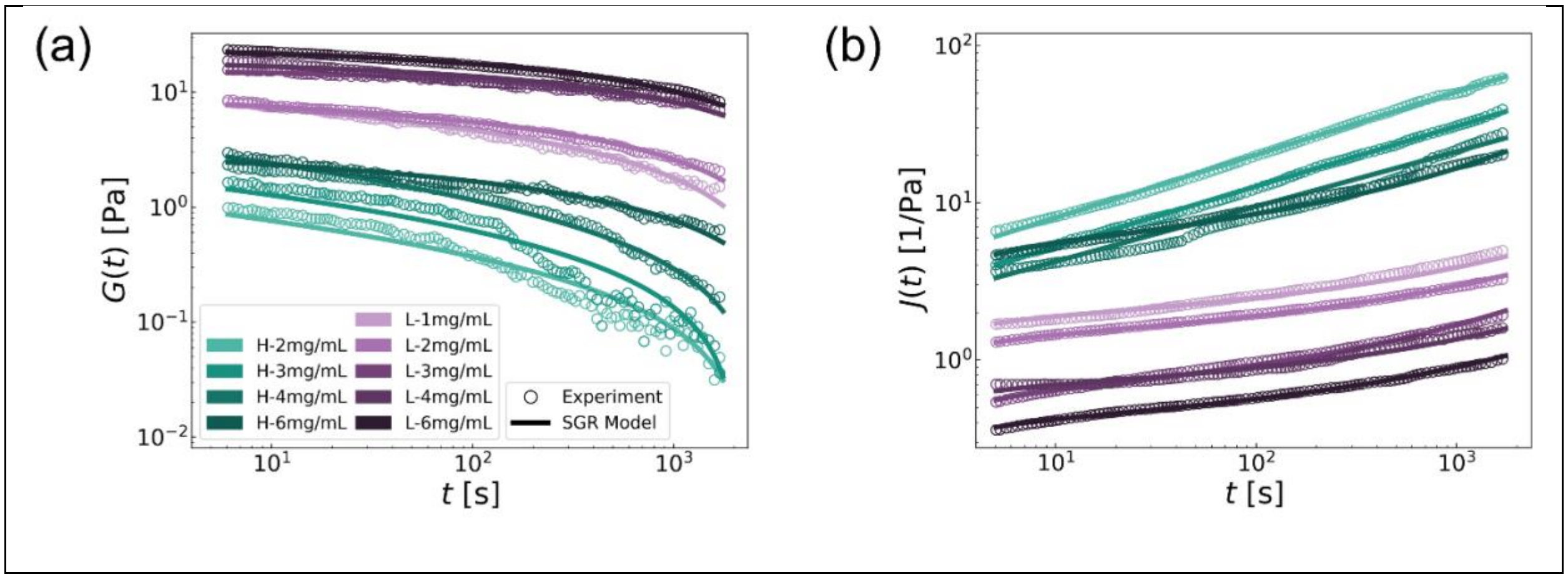

**Fig. 6** Long-time rheological response of high-brush (H) and low-brush (L) gels across PAM concentrations. **(a)** Stress-relaxation modulus following a 0.5% step strain and **(b)** creep compliance under a constant stress of 0.0105 Pa. Samples were aged for 7200 s prior to measurement, and each measurement was performed for 1800 s. Solid lines represent SGR fits. Low-brush gels retain higher relaxation moduli and exhibit lower creep compliance than high-brush gels, consistent with slower relaxation and deeper dynamic arrest.

To quantify these time-dependent rheological responses, we extract the effective noise temperature, $x$, from the SGR fits. Within the SGR framework, lower values of $x$ correspond to more deeply arrested dynamics, whereas larger values indicate a greater propensity for rearrangement within the colloidal network and a more fluid-like response. For stress relaxation, $x$ is obtained by fitting the relaxation modulus for each sample. For creep, the appropriate SGR regime is selected based on the relaxation-derived value of $x$, and the creep response is then fitted independently to obtain a creep-derived value of $x$. Figure 7 plots the resulting effective noise temperatures extracted from both stress-relaxation and creep measurements as functions of depletant concentration for the high-brush and low-brush gels. This comparison allows the degree of dynamic arrest to be quantified across brush density, depletant concentration, and rheological protocol.

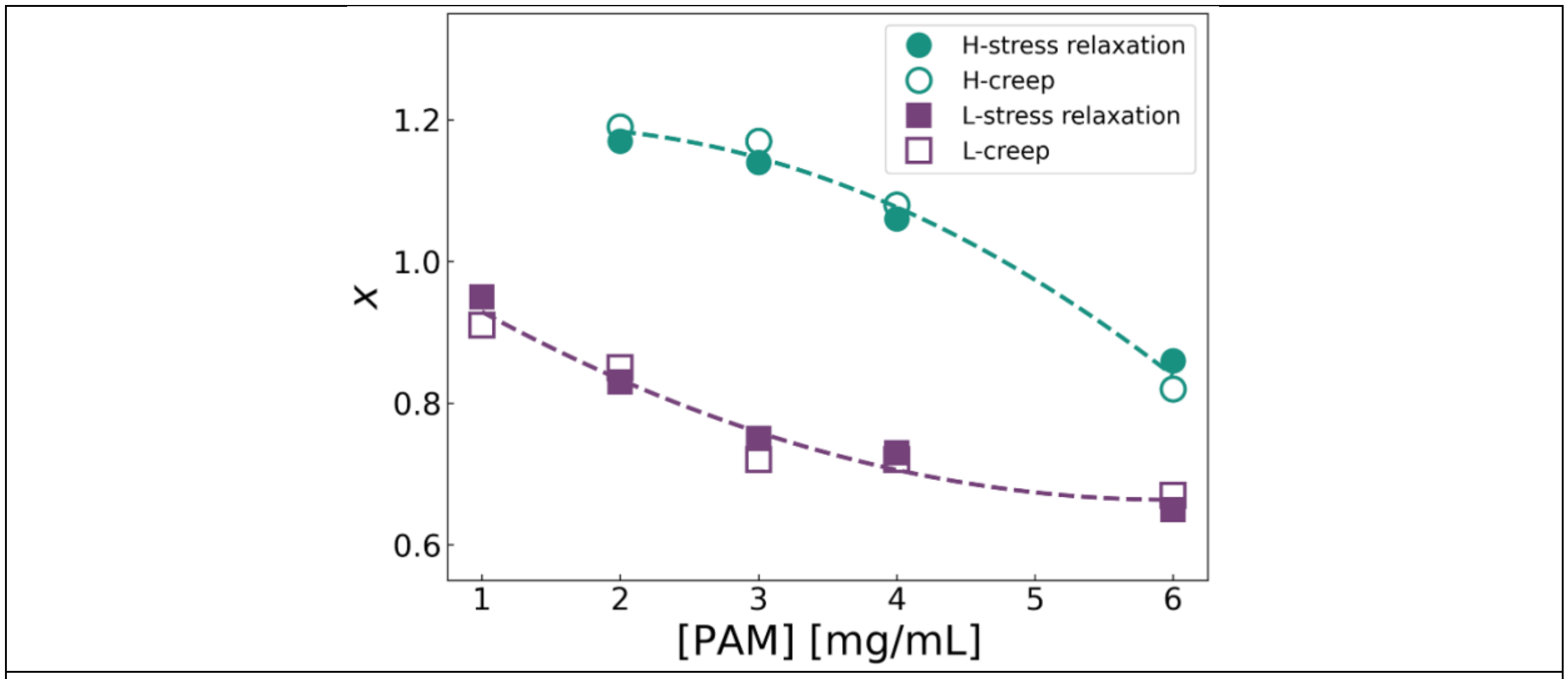

**Fig. 7** Effective noise temperature, $x$, extracted from SGR fits to stress-relaxation and creep measurements for high-brush (H) and low-brush (L) gels as functions of PAM concentration, [PAM]. Creep-derived values of $x$ were obtained independently using the SGR regime selected from the relaxation-derived $x$. Both protocols show lower $x$ values for low-brush gels, indicating more deeply arrested dynamics. Dashed lines are guides to the eye.

The extracted values of $x$ are consistent between the two rheological protocols, and follow the same overall trends. For both brush densities, $x$ decreases with increasing depletant concentration, consistent with stronger interparticle attraction resulting in more arrested dynamics. At matched depletant concentration, low-brush gels exhibit lower values of $x$ than high-brush gels, indicating that reduced brush density is associated with more deeply arrested dynamics. The agreement between relaxation- and creep-derived trends provides an additional check on the applicability of the SGR framework to these gels, showing that two independent rheological protocols give consistent trends in the extracted effective noise temperature. Modest differences between relaxation- and creep-derived $x$ values are observed near the crossover between more fluid-like and glassy regimes, where the distinction between SGR asymptotic forms becomes less sharp and the finite experimental time window can influence the fitted response.

Together, the relaxation and creep measurements show that reducing brush density affects not only the elastic and yielding response of the gels, but also their long-time rheological dynamics. Low-brush gels relax stress more slowly, accumulate less creep deformation, and exhibit lower effective noise temperatures across both rheological protocols. These results indicate that the enhanced load-bearing response of the low-brush gels is accompanied by deeper dynamic arrest.

## Conclusion

In this work, we investigated how reducing surface brush density affects the formation, structure, mechanics, and long-time dynamics of depletion-induced colloidal gels. By comparing high-brush and low-brush particles that remain geometrically indistinguishable in size and shape, we examined how surface-brush architecture alters rheology without introducing particle geometry as an additional variable.

Reducing brush density accelerates gelation and produces gels with higher plateau storage modulus and yield stress. These changes are not accompanied by increased local densification. Instead, low-brush networks exhibit lower average contact number and reduced spatial heterogeneity while still displaying stronger elastic and yielding responses than the corresponding high-brush gels. This comparison indicates that the rheological enhancement cannot be understood simply as a consequence of forming more contacts or denser local packing.

Coordination-based network analysis further clarifies this distinction. When the classical central-force isostatic threshold is used as a proxy for identifying load-bearing particles, the high-brush system contains the larger load-bearing particle fraction, opposite to the observed rheological trend. In contrast, applying a reduced coordination criterion motivated by brush-mediated resistance to angular deformation gives a larger load-bearing particle fraction for the low-brush system, consistent with its higher stiffness and yield stress. This contrast suggests that reducing brush density allows angular resistance at particle contacts to contribute to stress support, enabling lower-coordination networks to support load more effectively than expected from the classical central-force isostatic criterion.

The same reduction in brush density also affects long-time rheological properties beyond linear viscoelasticity. Low-brush gels retain higher relaxation moduli, accumulate less creep deformation, and exhibit lower effective noise temperatures than high-brush gels across both stress-relaxation and creep measurements. The consistent trends obtained from these two rheological protocols provide an additional check on the applicability of the Soft Glassy Rheology framework to these gels and indicate more deeply arrested dynamics in the low-brush systems.

Together, these findings establish surface-brush density as a geometry-preserving control parameter for gelation kinetics, oscillatory mechanics, and long-time linear rheology in colloidal depletion gels. The combination of enhanced stiffness, reduced coordination, and slower normalized relaxation is consistent with a picture in which lower brush density modifies both the effective attraction and the kinematic constraints at particle contacts. The present coordination analysis supports, but does not independently prove, brush-mediated angular resistance or a unique reduced rigidity threshold. More broadly, the results demonstrate that surface architecture can tune gel rheology without requiring changes in primary particle shape or roughness.

## Declarations

**Competing interests:** The authors declare no competing interests.

**Funding:** This work was supported by the National Science Foundation under Grant No. PMP-2025613. Z.Z. and A.M. acknowledge support from the UC Irvine Materials Research Institute, which is supported in part by the National Science Foundation through the UC Irvine Materials Research Science and Engineering Center under Grant No. DMR-2011967.

**Data availability:** The datasets generated during and/or analyzed during the current study are available from the corresponding author on reasonable request.

## References

1. Akpinar B, Fielding LA, Cunningham VJ, Ning Y, Mykhaylyk OO, Fowler PW, Armes SP (2016) Determining the Effective Density and Stabilizer Layer Thickness of Sterically Stabilized Nanoparticles. Macromolecules, 49(14), 5160–5171. https://doi.org/10.1021/acs.macromol.6b00987
2. Allan DB, Caswell T, Keim NC, van der Wel, CM, Verweij RW (2025) soft-matter/trackpy: V0.7 (Version v0.7) [Computer software]. Zenodo. https://doi.org/10.5281/ZENODO.16089574

3. Asakura S, Oosawa F (1958) Interaction between particles suspended in solutions of macromolecules. Journal of Polymer Science, 33(126), 183–192. https://doi.org/10.1002/pol.1958.1203312618
4. Bianchi E, Largo J, Tartaglia P, Zaccarelli E, Sciortino F (2006) Phase Diagram of Patchy Colloids: Towards Empty Liquids. Physical Review Letters, 97(16), 168301. https://doi.org/10.1103/PhysRevLett.97.168301
5. Buscall R, Mills PDA, Goodwin JW, Lawson DW (1988) Scaling behaviour of the rheology of aggregate networks formed from colloidal particles. Journal of the Chemical Society, Faraday Transactions 1: Physical Chemistry in Condensed Phases, 84(12), 4249. https://doi.org/10.1039/f19888404249
6. Cipelletti L, Manley S, Ball RC, Weitz DA (2000) Universal Aging Features in the Restructuring of Fractal Colloidal Gels. Physical Review Letters, 84(10), 2275–2278. https://doi.org/10.1103/PhysRevLett.84.2275
7. Colombo J, Del Gado E (2014) Stress localization, stiffening, and yielding in a model colloidal gel. Journal of Rheology, 58(5), 1089–1116. https://doi.org/10.1122/1.4882021
8. Del Gado E, Kob W (2010) A microscopic model for colloidal gels with directional effective interactions: Network induced glassy dynamics. Soft Matter, 6(7), 1547. https://doi.org/10.1039/b916813c
9. Dias CS, Neves JC, Telo Da Gama MM, Del Gado E, Araújo NAM (2025) Structural criterion for the onset of rigidity in a colloidal gel. Physical Review E, 112(3), L033401. https://doi.org/10.1103/gw42-dks6
10. Fielding SM, Sollich P, Cates ME (2000) Aging and rheology in soft materials. Journal of Rheology, 44(2), 323–369. https://doi.org/10.1122/1.551088
11. Hsiao LC, Newman RS, Glotzer SC, Solomon MJ (2012) Role of isostaticity and load-bearing microstructure in the elasticity of yielded colloidal gels. Proceedings of the National Academy of Sciences, 109(40), 16029–16034. https://doi.org/10.1073/pnas.1206742109
12. Hsiao LC, Solomon MJ., Whitaker KA, Furst EM (2014) A model colloidal gel for coordinated measurements of force, structure, and rheology. Journal of Rheology, 58(5), 1485–1504. https://doi.org/10.1122/1.4884965
13. Hsiao LC, Jamali S, Glynos E, Green PF, Larson RG, Solomon MJ (2017) Rheological State Diagrams for Rough Colloids in Shear Flow. Physical Review Letters, 119(15), 158001. https://doi.org/10.1103/PhysRevLett.119.158001
14. Hsu CP, Ramakrishna SN, Zanini M, Spencer ND, Isa L (2018) Roughness-dependent tribology effects on discontinuous shear thickening. Proceedings of the National Academy of Sciences, 115(20), 5117–5122. https://doi.org/10.1073/pnas.1801066115
15. Huh JY, Lynch ML, Furst EM (2007) Microscopic structure and collapse of depletion-induced gels in vesicle-polymer mixtures. Physical Review E, 76(5), 051409. https://doi.org/10.1103/PhysRevE.76.051409

16. Ilett SM, Orrock A, Poon WCK, Pusey PN (1995) Phase behavior of a model colloid-polymer mixture. Physical Review E, 51(2), 1344–1352. https://doi.org/10.1103/PhysRevE.51.1344
17. Immink JN, Maris JJE, Schurtenberger P, Stenhammar J (2020) Using Patchy Particles to Prevent Local Rearrangements in Models of Non-equilibrium Colloidal Gels. Langmuir, 36(1), 419–425. https://doi.org/10.1021/acs.langmuir.9b02675
18. Johnson LC, Zia RN, Moghimi E, Petekidis G (2019) Influence of structure on the linear response rheology of colloidal gels. Journal of Rheology, 63(4), 583–608. https://doi.org/10.1122/1.5082796
19. Kim J, Merger D, Wilhelm M, Helgeson ME (2014) Microstructure and nonlinear signatures of yielding in a heterogeneous colloidal gel under large amplitude oscillatory shear. Journal of Rheology, 58(5), 1359–1390. https://doi.org/10.1122/1.4882019
20. Kim S, Hyun K, Moon JY, Clasen C, Ahn KH (2015) Depletion Stabilization in Nanoparticle–Polymer Suspensions: Multi-Length-Scale Analysis of Microstructure. Langmuir, 31(6), 1892–1900. https://doi.org/10.1021/la504578x
21. Kobelev V, Schweizer KS (2005) Nonlinear elasticity and yielding of depletion gels. J. Chem. Phys. 123, 164902. https://doi.org/10.1063/1.2109887
22. Kodger TE, Guerra RE, Sprakel J (2015) Precise colloids with tunable interactions for confocal microscopy. Scientific Reports, 5(1), 14635. https://doi.org/10.1038/srep14635
23. Krall AH, Weitz DA (1998) Internal Dynamics and Elasticity of Fractal Colloidal Gels. Physical Review Letters, 80(4), 778–781. https://doi.org/10.1103/PhysRevLett.80.778
24. Landrum BJ, Russel WB, Zia RN (2016) Delayed yield in colloidal gels: Creep, flow, and re-entrant solid regimes. Journal of Rheology, 60(4), 783–807. https://doi.org/10.1122/1.4954640
25. Laurati M, Egelhaaf SU, Petekidis G (2011) Nonlinear rheology of colloidal gels with intermediate volume fraction. Journal of Rheology, 55(3), 673–706. https://doi.org/10.1122/1.3571554
26. Louis AA, Bolhuis PG, Meijer EJ, Hansen JP (2002) Polymer induced depletion potentials in polymer-colloid mixtures. The Journal of Chemical Physics, 117(4), 1893–1907. https://doi.org/10.1063/1.1483299
27. Lu PJ, Zaccarelli E, Ciulla F, Schofield AB, Sciortino F, Weitz DA (2008) Gelation of particles with short-range attraction. Nature, 453(7194), 499–503. https://doi.org/10.1038/nature06931
28. Mangal D, Nabizadeh M, Jamali S (2023) Topological origins of yielding in short-ranged weakly attractive colloidal gels. The Journal of Chemical Physics, 158(1), 014903. https://doi.org/10.1063/5.0123096
29. Mangal D, Nabizadeh M, Jamali S (2024) Predicting yielding in attractive colloidal gels. Physical Review E, 109(1), 014602. https://doi.org/10.1103/PhysRevE.109.014602

30. Maxwell JC (1864) L. On the calculation of the equilibrium and stiffness of frames. The London, Edinburgh, and Dublin Philosophical Magazine and Journal of Science, 27(182), 294–299. https://doi.org/10.1080/14786446408643668

31. Mewis J, Wagner NJ (2013) Colloidal suspension rheology (1. paperback ed). Cambridge University Press.

32. Müller FJ, Isa L, Vermant J (2023) Toughening colloidal gels using rough building blocks. Nature Communications, 14(1), 5309. https://doi.org/10.1038/s41467-023-41098-9

33. Nabizadeh M, Nasirian F, Li X, Saraswat Y, Waheibi R, Hsiao LC, Bi D, Ravandi B, Jamali S (2024) Network physics of attractive colloidal gels: Resilience, rigidity, and phase diagram. Proceedings of the National Academy of Sciences, 121(3), e2316394121. https://doi.org/10.1073/pnas.2316394121

34. Nguyen HT, Graham AL, Koenig PH, Gelb LD (2020) Computer simulations of colloidal gels: How hindered particle rotation affects structure and rheology. Soft Matter, 16(1), 256–269. https://doi.org/10.1039/C9SM01755K

35. Palombo G, Weir S, Michieletto D, Gutiérrez Fosado YA (2025) Topological linking determines elasticity in limited valence networks. Nature Materials, 24(3), 454–461. https://doi.org/10.1038/s41563-024-02091-9

36. Pantina JP, Furst EM (2005) Elasticity and Critical Bending Moment of Model Colloidal Aggregates. Physical Review Letters, 94(13), 138301. https://doi.org/10.1103/PhysRevLett.94.138301

37. Park N, Umanzor EJ, Conrad JC (2018) Aqueous Colloid + Polymer Depletion System for Confocal Microscopy and Rheology. Frontiers in Physics, 6, 42. https://doi.org/10.3389/fphy.2018.00042

38. Pincus P (1991) Colloid stabilization with grafted polyelectrolytes. Macromolecules, 24(10), 2912–2919. https://doi.org/10.1021/ma00010a043

39. Radhakrishnan R, Divoux T, Manneville S, Fielding SM (2017) Understanding rheological hysteresis in soft glassy materials. Soft Matter, 13(9), 1834–1852. https://doi.org/10.1039/C6SM02581A

40. Rajaram B, Mohraz A(2010) Microstructural response of dilute colloidal gels to nonlinear shear deformation. Soft Matter, 6(10), 2246. https://doi.org/10.1039/b926076e

41. Ramakrishnan S, Chen YL, Schweizer KS, Zukoski CF (2004) Elasticity and clustering in concentrated depletion gels. Physical Review E, 70(4), 040401. https://doi.org/10.1103/PhysRevE.70.040401

42. Rocklin DZ, Hsiao L, Szakasits M, Solomon MJ, Mao X (2021) Elasticity of colloidal gels: Structural heterogeneity, floppy modes, and rigidity. Soft Matter, 17(29), 6929–6934. https://doi.org/10.1039/D0SM00053A

43. Scherrer S, Ramakrishna SN, Niggel V, Hsu CP, Style RW, Spencer ND, Isa L (2025) Characterizing sliding and rolling contacts between single particles. Proceedings of the

National Academy of Sciences, 122(10), e2411414122. https://doi.org/10.1073/pnas.2411414122

44. Schweizer KS, Saltzman EJ (2003) Entropic barriers, activated hopping, and the glass transition in colloidal suspensions. J. Chem. Phys. 119, 1181–1196. https://doi.org/10.1063/1.1578632
45. Sciortino F, Zaccarelli E (2017) Equilibrium gels of limited valence colloids. Current Opinion in Colloid & Interface Science, 30, 90–96. https://doi.org/10.1016/j.cocis.2017.06.001
46. Sollich P, Lequeux F, Hébraud P, Cates ME (1997) Rheology of Soft Glassy Materials. Physical Review Letters, 78(10), 2020–2023. https://doi.org/10.1103/PhysRevLett.78.2020
47. Sprakel J, Lindström SB., Kodger TE, Weitz DA (2011) Stress Enhancement in the Delayed Yielding of Colloidal Gels. Physical Review Letters, 106(24), 248303. https://doi.org/10.1103/PhysRevLett.106.248303
48. Studart AR, Amstad E, Gauckler LJ (2011) Yielding of weakly attractive nanoparticle networks. Soft Matter, 7(14), 6408. https://doi.org/10.1039/c1sm05598d
49. Trappe V, Prasad V, Cipelletti L, Segre PN, Weitz, D A (2001) Jamming phase diagram for attractive particles. Nature, 411(6839), 772–775. https://doi.org/10.1038/35081021
50. Varga Z, Swan JW (2015) Linear viscoelasticity of attractive colloidal dispersions. Journal of Rheology, 59(5), 1271–1298. https://doi.org/10.1122/1.4928951
51. Wang G, Swan JW (2019) Surface heterogeneity affects percolation and gelation of colloids: Dynamic simulations with random patchy spheres. Soft Matter, 15(25), 5094–5108. https://doi.org/10.1039/C9SM00607A
52. Weitz DA, Oliveria M (1984) Fractal Structures Formed by Kinetic Aggregation of Aqueous Gold Colloids. Physical Review Letters, 52(16), 1433–1436. https://doi.org/10.1103/PhysRevLett.52.1433
53. Whitaker KA, Varga Z, Hsiao LC, Solomon MJ, Swan JW, Furst EM (2019) Colloidal gel elasticity arises from the packing of locally glassy clusters. Nature Communications, 10(1), 2237. https://doi.org/10.1038/s41467-019-10039-w
54. Wolthers W, Van Den Ende D, Breedveld V, Duits MHG, Potanin AA, Wientjes RHW, Mellema J (1997) Linear viscoelastic behavior of aggregated colloidal dispersions. Physical Review E, 56(5), 5726–5733. https://doi.org/10.1103/PhysRevE.56.5726
55. Yanez JA, Laarz E, Bergström L (1999) Viscoelastic Properties of Particle Gels. Journal of Colloid and Interface Science, 209(1), 162–172. https://doi.org/10.1006/jcis.1998.5892
56. Yang S, Yan D, Tan H, Shi AC (2006) Depletion interaction between two colloidal particles in a nonadsorbing polymer solution. Physical Review E, 74(4), 041808. https://doi.org/10.1103/PhysRevE.74.041808

57. Yin G, Solomon MJ (2008) Soft glassy rheology model applied to stress relaxation of a thermoreversible colloidal gel. Journal of Rheology, 52(3), 785–800. https://doi.org/10.1122/1.2885738

58. Zaccarelli E (2007) Colloidal gels: Equilibrium and non-equilibrium routes. Journal of Physics: Condensed Matter, 19(32), 323101. https://doi.org/10.1088/0953-8984/19/32/323101

59. Zaccone A, Wu H, Del Gado E (2009) Elasticity of Arrested Short-Ranged Attractive Colloids: Homogeneous and Heterogeneous Glasses. Physical Review Letters, 103(20), 208301. https://doi.org/10.1103/PhysRevLett.103.208301

60. Zhuang Z, Campbell RA, Haghighi P, Jamali S, Mohraz A (2026) Brush-mediated angular constraints reshape structure, rigidity, and percolation in colloidal depletion gels. Under Publication. https://doi.org/10.48550/arXiv.2603.13596

# Supplementary Information:

# Rheology and Dynamic Arrest in Colloidal Depletion Gels Mediated by Surface Brush Density

Ziye Zhuang[1], Robert A. Campbell[2], Safa Jamali[2,3], and Ali Mohraz[1*]

*1. Department of Chemical and Biomolecular Engineering, University of California, Irvine, Irvine, CA 92697, USA*

*2. Department of Mechanical and Industrial Engineering, Northeastern University, Boston, MA 02115, USA*

*3. Department of Chemical Engineering, Northeastern University, Boston, MA 02115, USA*

**Corresponding author: mohraz@uci.edu*

## S1. Colloidal stability in the absence of PAM

To determine whether the salt concentration used in this study could independently induce particle aggregation, control suspensions of high- and low-brush particles were prepared at a particle volume fraction of $\phi = 0.20$ and a NaCl concentration of 10 mM in the absence of PAM. Representative confocal snapshots of both suspensions in Figure S1 show no obvious aggregation or network formation, indicating that the gelation observed in the main text is primarily driven by depletion attraction rather than salt alone.

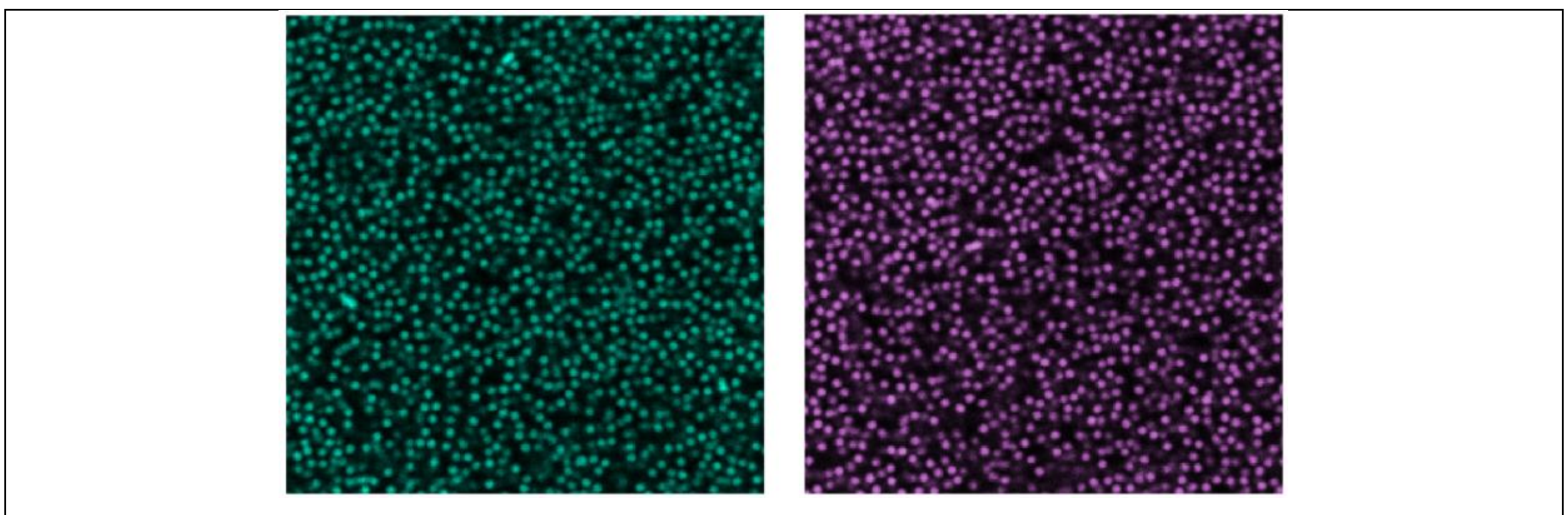

**Fig. S1** Representative confocal snapshots of high-brush (left) and low-brush (right) suspensions at a particle volume fraction of $\phi = 0.20$ and a NaCl concentration of 10 mM in the absence of PAM. No obvious aggregation or network formation is observed.

## S2. Supplementary methodological details

### S2.1 Justification of particle neighbor search distance

To quantify gel microstructure from confocal image stacks, we identified particle neighbors based on a surface-to-surface search distance of 0.4 μm. Although this distance is relatively large compared to the particle diameter (1.06 μm), it accounts for experimental limitations such as particle motion during acquisition, limited spatial resolution especially in the z direction, and microscope imaging noise, all of which can compromise reliable contact detection if a stricter criterion is used. This threshold provided robust coordination number statistics for all samples.

### S2.2 Explicit expressions used in SGR creep analysis

To implement the numerical solution of the SGR creep equation described in the main text, we use the following expressions for the yielding rate $Y(t)$ and the survival function $G_\rho(t - t')$ provided by Fielding and coworkers (Fielding et al. 2000).

**Yielding rate $\boldsymbol{Y(t)}$.** For the glassy regime ($x < 1$), the yielding rate is given by the asymptotic form (Fielding et al. 2000, Eq. 37):

$$Y(t) = \frac{t^{x-1}}{x\,\Gamma(x)\Gamma(1-x)}$$

where $\Gamma(\cdot)$ is the gamma function and $x$ is the effective noise temperature.

**Survival function $\boldsymbol{G_\rho(t - t')}$.** The survival function is expressed as (Fielding et al. 2000, Eq. 28):

$$G_\rho(t - t') = x \int_1^\infty \tau^{-x-1} \exp\left(-\frac{t - t'}{\tau}\right) d\tau$$

This function represents the retained elastic contribution from elements that remain trapped between times $t'$ and $t$.

## S3. Extended time-sweep of the 1 mg/mL high-brush sample

To determine whether the high-brush suspension at 1 mg/mL PAM is capable of forming a load-bearing network within experimentally relevant timescales, we performed an extended small-amplitude oscillatory shear (SAOS) time sweep using the same conditions described in the main text (0.5% strain, 1 rad/s, pre-sheared), but with the measurement window increased from the standard 2 h to 4 h as shown in Figure S2.

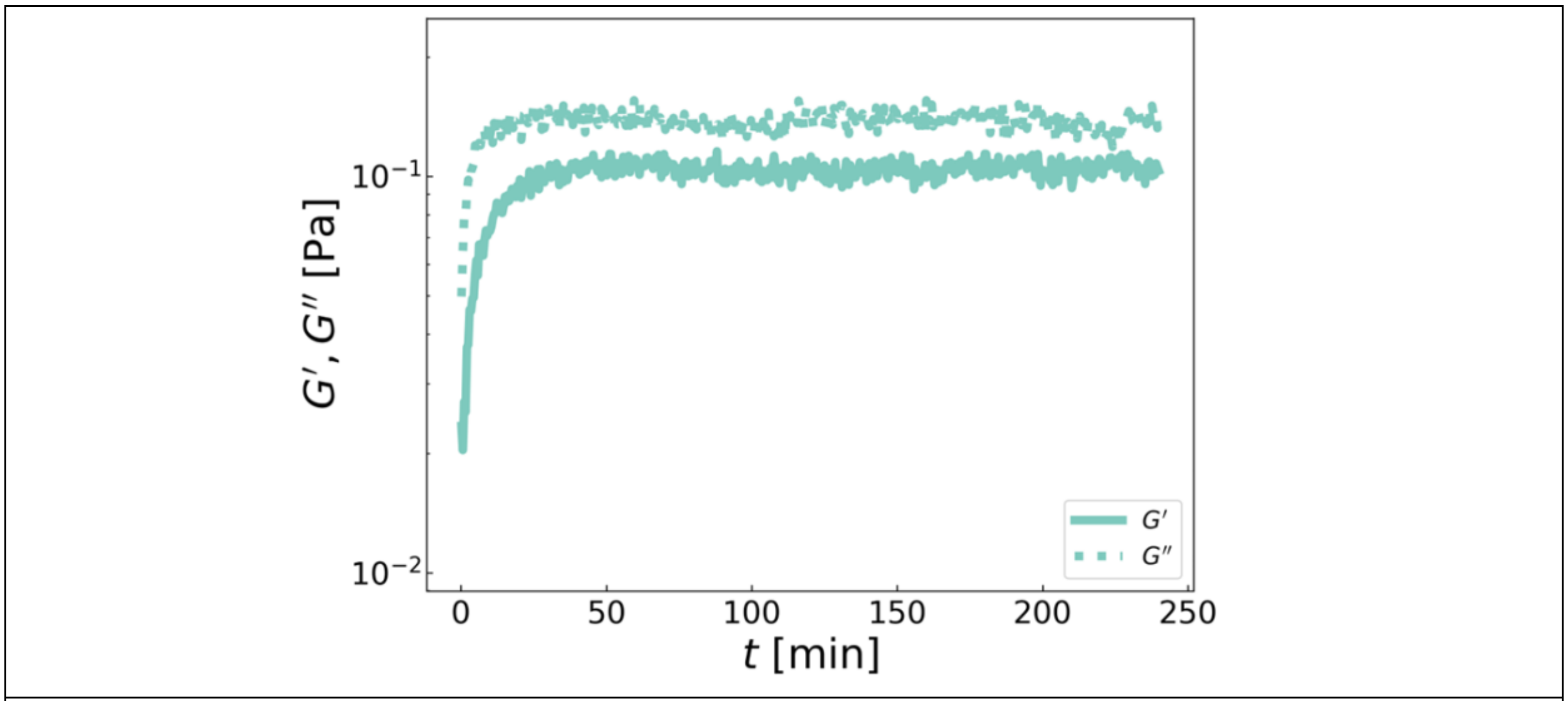

**Fig. S2** Extended time-sweep (4 h) of the high-brush 1 mg/mL sample showing that $G'$ remains below $G''$ for the full duration, indicating no gelation is observed under these conditions.

Throughout the entire 4 h test, the storage modulus remained below the loss modulus, $G' < G''$, confirming the absence of gelation. Although $G'$ exhibited an initial rise and reached a shallow plateau at long times, this growth was insufficient to overcome $G''$, indicating that the system remained fluid-like. This result verifies that 1 mg/mL PAM does not provide adequate depletion attraction for gelation in the high-brush system, even under substantially extended waiting times. For this reason, the high-brush 1 mg/mL condition was excluded from subsequent rheological analyses.

## S4. Additional structural analysis

### S4.1 Network connectivity (Largest Connected Component)

To verify that gelled samples form system-spanning networks, we computed the fraction of particles belonging to the largest connected component (LCC) for each condition, as shown in Figure S3. For all samples with PAM concentrations ≥ 2 mg/mL, both high-brush and low-brush systems exhibit near-unity LCC fractions, indicating that most particles are incorporated into a single percolated network. In contrast, the high-brush sample at 1 mg/mL PAM shows a substantially lower LCC fraction, consistent with its failure to develop a solid-like viscoelastic response within the experimental window (Figure S2). These results show that, for the gelled conditions analyzed in the main text, differences in stiffness and yielding are not due to incomplete percolation, but instead are associated with differences in network structure and brush-mediated contact mechanics.

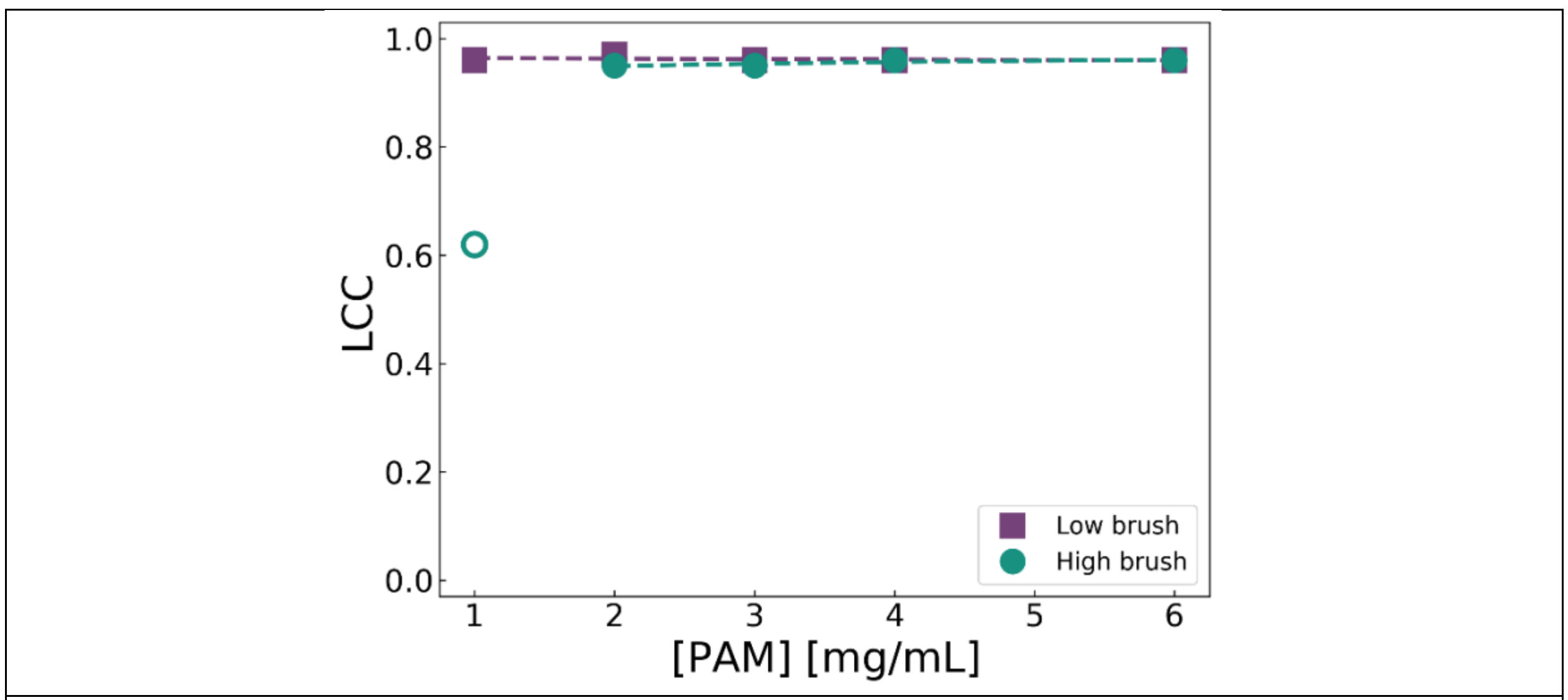

**Fig. S3** Fraction of particles belonging to the largest connected component (LCC) for high-brush (H) and low-brush (L) systems as functions of PAM concentration, [PAM]. All gelled samples (≥2 mg/mL PAM) exhibit near-unity connectivity, indicating the formation of system-spanning networks.

### S4.2 Full distributions of contact number and Voronoi volume

Figure S4 shows the complete contact number distributions and Voronoi volume distributions for all polymer concentrations. These plots accompany the representative distributions shown in the main text (Figure 4a,b) and show the full range of structural variation across conditions.

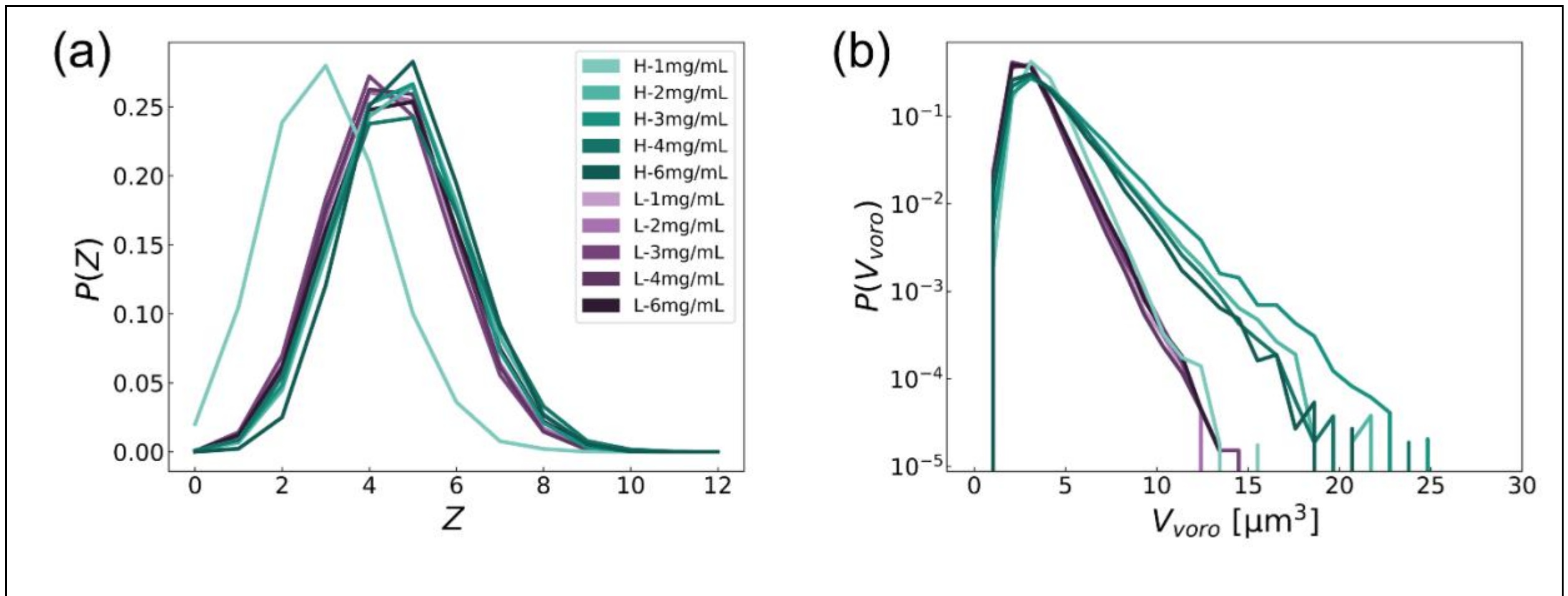

**Fig. S4 (a)** Full particle contact number distributions for high-brush (H) and low-brush (L) samples across all PAM concentrations (1–6 mg/mL). **(b)** Full Voronoi volume distributions for the same set of samples.